\documentclass[a4paper,10pt]{extarticle}

\usepackage{epsfig}
\usepackage{longtable}
\usepackage{fancyhdr}
\usepackage{hhline}
\usepackage{array}
\usepackage{rotating}
\usepackage{cite}
\usepackage{amsfonts}
\usepackage{amssymb}
\usepackage{amsmath}
\usepackage{underscore}
\usepackage{color}
\usepackage{afterpage}
\usepackage[latin1]{inputenc}
\usepackage{geometry}
\usepackage{fancybox}
\usepackage{bibtopic}
\usepackage{relsize}
\usepackage{calc}
\usepackage[a4paper,ps2pdf,colorlinks=true,linkcolor=red,urlcolor=blue,breaklinks=true]{hyperref}
\usepackage[yyyymmdd]{datetime}
\usepackage{multicol}
\usepackage{enumitem}


\newcommand{\myses}[2]{\subsection*{\raisebox{-0.3mm}{\rule{2mm}{3mm}}\hspace{7mm}{\textbf{#1}}} 
\vspace{-2mm} \hspace{9mm}\textit{Chairperson:} #2 \vspace{1mm}}

\newcommand{\mysec}[2]{\subsection*{\raisebox{-0.3mm}{\rule{2mm}{3mm}}\hspace{7mm}{\textbf{#1}}}}

\newlength{\defitemindent} 

\newenvironment{mylist}[1]%
{
\begin{list}{}
   {\itemsep=2pt \parsep=0pt \topsep=0pt \parskip=1pt \rightmargin=0mm
   \settowidth{\labelwidth}{\hspace{\defitemindent}\bf #1}%
   \setlength{\leftmargin}{\labelwidth}%
   \addtolength{\leftmargin}{\labelsep}%
   }}%
{\vspace{-2mm}\end{list}
}

{
\begin{list}{}
   {\itemsep=0pt \parsep=0pt \topsep=0pt \parskip=1pt \rightmargin=0mm
   \settowidth{\labelwidth}{\hspace{\defitemindent}\bf #1}%
   \setlength{\leftmargin}{\labelwidth}%
   \addtolength{\leftmargin}{\labelsep}%
   }}%
{\vspace{1mm}\end{list}
}

\allowdisplaybreaks

\usepackage{charter}

\newcommand{\hide}[1]{}

\newcommand{\conta}[5]{\item[#1] \textsc{#2} \textit{#3} \\ {#4} \\[1mm]
   {\small #5}}

\newcommand{\cont}[4]{\item[#1] \textsc{#2} \textit{#3} \\ {#4}}

\begin{document}

\thispagestyle{empty}

\noindent
\raisebox{-24.5mm}{\rule{2mm}{50mm}} 
\hspace{5mm}
\begin{minipage}{\textwidth/\real{1.2}}
Proceedings of the \\[3mm]
\hspace*{0.3mm}{\Large\textsc{\textbf{2013 UK MHD --}} \\[3mm] 
National Conference on 
Geophysical, Astrophysical  and Industrial \\[0.5mm] Magnetohydrodynamics} \\[5mm]
{\large Glasgow, 23 and 24 May 2013} \\[5mm]  
\epsfig{file=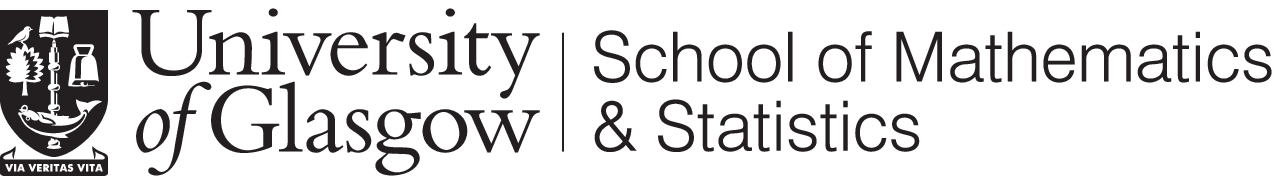,height=11mm,clip=}
\raisebox{2mm}{\rule{0.1mm}{7mm}}  
\raisebox{5mm}{ \begin{minipage}{4cm} 
{\footnotesize \textbf{Radostin D.~Simitev}\\[-1.5mm] Andrew  W.~Baggaley \\[-1.5mm] David R.~Fearn}
\end{minipage}}
\end{minipage}

\section*{--- Editorial note ---}{}

The UK National Conference on Geophysical, Astrophysical and
Industrial Magnetohydrodynamics (UK MHD) is an annual meeting of the
UK Magnetohydrodynamics research community. The meeting has been
organized every year since 1978 with the exception of 1983 and 1987, 
(\url{http://www.maths.gla.ac.uk/~drf/mhd.htm}). The 2013 meeting was
held at the School of Mathematics and Statistics, the University of
Glasgow on the 23 and 24 May. This document presents an accurate
record of the scientific programme of the meeting. The abstracts of
presented contributions are listed in chronological order and a full
list of participants and their affiliations is
included. 
The Meeting was sponsored by 
\begin{itemize}[itemsep=-1pt,parsep=2pt]
\item The London Mathematical Society, \url{www.lms.ac.uk}, 
\item The Glasgow Journal of Mathematics Trust, \url{www.maths.gla.ac.uk/gmj/gmj-trust},
\item The Edinburgh Mathematical Society \url{www.ems.ac.uk},
\item The Science and Technology Facilities Council \url{www.stfc.ac.uk}.
\end{itemize}
The conference website
(\url{http://www.gla.ac.uk/schools/mathematicsstatistics/events/ukmhd})
records some further details of the meeting.

\textsc{Radostin D.~Simitev}, \textit{A.W.~Baggaley \&  D.R.~Fearn}

\section*{--- Programme ---}{}

The original version this programme as used at the time of the Meeting
can be found online at\\
\url{http://www.gla.ac.uk/schools/mathematicsstatistics/events/ukmhd}.

\subsubsection*{--- Thursday, 2013-05-23 ---}{}

\vspace{3mm}
\mysec{Welcome and Registration}{}
\begin{mylist}{19:00 -- 10:30 am}
    \item[9:00 -- 10:25] Arrival, registration, set-up of
      posters. Refreshments available.
    \cont{10:25 -- 10:30}  {Prof N.A.~Hill}{}{Head of School Welcome message}
\end{mylist}

\myses{Invited Lecture}{David Fearn}
\begin{mylist}{19:00 -- 10:30 am}
      \conta{10:30 -- 11:10}
          {Andrew M.~Soward,}
          {A.P.~Bassom, K.~Kuzanyan, D.~Sokoloff and S.~Tobias} 
          {Asymptotic solution of a kinematic $\alpha\Omega$--dynamo
            with meridional circulation}
          {Many stars exhibit magnetic cycles typified by the
    butterfly diagram characterising our sun's 11 year solar activity
    cycle. Parker explained the phenomenon by an $\alpha\Omega$--dynamo acting in
    the star's convection zone causing the equatorial propagation of
    dynamo waves. In contrast, to the many continuing numerical
    investigations, we adopt a minimalist approach and expand on
    Parker's original one-dimensional uniform plane layer model. To
    apply asymptotic methods, we suppose that the dynamo is confined
    to a thin shell with latitudinal variations of the $\alpha\Omega$--sources,
    whose product the Dynamo number vanishes at the pole and
    equator. The ensuing linear stability problem is resolved by
    global stability criteria. Our new results concern the role of
    meridional circulation. They show that sufficiently large
    circulation halts the Parker travelling waves leading to
    non-oscillatory behaviour, a result only predicted previously from
    numerical integration of the full pde's governing axisymmetric
    $\alpha\Omega$--dynamos. 
}
\end{mylist}          

\myses{Session 1: Geophysical and Planetary Applications I}{Radostin Simitev}
\begin{mylist}{19:00 -- 10:30 am}
      \conta{11:10 -- 11:25} 
           {William Brown,}
           {J.~Mound, P.~Livermore}
           {Jerks abound: Observations of geomagnetic  jerks and
             implications for core dynamics}
           {The geomagnetic field is generated by the constant
    evolution of the fluid outer core.  The secular variation of the
    geomagnetic field describes the variation of the field at time
    scales on the order of months to decades and is attributed
    primarily to flows near the surface of the outer core.  The
    secular variation, at any point on the Earth's surface, is often
    characterised as a series of linear trends separated by abrupt
    turning points known as geomagnetic jerks.  These rapid variations
    in the secular variation are linked to accelerations of flow in
    the outer core.  Various generation mechanisms for these rapid
    changes have been suggested but none have conclusively explained
    the phenomena.  Our recent study of geomagnetic jerks in
    observatory data over the period of 1957-2008 indicates that jerks
    are far more frequent an occurrence than previously suggested and
    perhaps part of the more rapid end of a spectrum of core
    dynamics.  Whilst jerks are seen to be common, relative peaks in the global number of jerk occurrences are
 seen in 1968-71, 1973-74, 1977-79, 1983-85, 1989-93, 1995-98 and
 2002-03 with the suggestion of further poorly sampled events in the
 early 1960s and late 2000s.  We do not find consistent patterns in
 the spatial distributions of occurrences suggesting complex origins
 or the superposition of several discrete individual events. We
 observe that jerk amplitudes vary through time and their variations
 are potentially periodic in Europe and North America.  This may have
 implications which help to constrain a source mechanism in the
 dynamics of the outer core.  These signals may be related to the 6yr
 periods detected independently in the secular variation and
 length-of-day.
}
      \conta{11:25 -- 11:40}
           {Robert Teed,}
           {C.~Jones and S.~Tobias}
           {Torsional oscillations in geodynamo   simulations}
{
Torsional oscillations are a principal feature of
    the dynamics of the fluid outer core where the Earth's magnetic
    field is generated. These oscillations are Alfvén waves about an
    equilibrium known as a Taylor state and propagate in the
    cylindrical radial direction. The change in core angular momentum
    inferred from geomagnetic observations has a measurable impact on
    the length of the day, and the small decadal variations in the
    length-of-day signal confirm the existence of torsional
    oscillations.
In our work we perform three-dimensional spherical dynamo simulations
in parameter regimes where Earth-like magnetic fields are
produced. Many of our simulations frequently produce the desired
torsional oscillations, identified by their movement at the correct
Alfvén speed. We find that the frequency, location and direction of
propagation of the waves can be influenced by the choice of parameter
regime and boundary conditions. Torsional waves are observed within
the tangent cylinder region and also have the ability to pass through
this theoretical cylinder. Our results indicate that excitation
mechanisms for these waves must be available throughout the outer
core. We calculate the driving terms in an attempt to better
understand these mechanisms.}
      \conta{11:40 -- 11:55}
           {Luis~Silva,}
           {C.~Davies, J.~Mound}
           {Inner core structure inferred from
    observations of the geomagnetic field}
{The short term evolution of the geomagnetic field,
    as observed at Earth's surface, contains the information required
    to infer the velocity field at
the top of the core. Although this velocity field cannot be uniquely
estimated, several studies have demonstrated that some of its features
are robust, not only
across assumptions, but also over time. Due to strong rotational
effects, the geomagnetic field and the associated core-surface
velocities are expected to reflect the
dynamics of the bulk of the core, an assumption that has seldom been
used.
Here we use a very simplified model of Earth's core and make
inferences about its thermal structure starting from observationally
constrained core-surface flows.
Despite its simplicity, this model is in agreement with much more
elaborate data assimilation models making it useful to test
assumptions about the force balances
inside Earth's core. Finally, we relate the obtained thermal structure
of the outer core to the seismically observed structure of the inner
core an propose a possible
causal relationship.
}
      \conta{11:55 -- 12:10}
           {Wieland Dietrich,}
           {T.~Gastine and J.~Wicht}
           {The nonlinear interaction between
    geostrophic and ageostrophic zonal flows}
{Zonal flows on rapidly rotating objects, such as the
    Gas or the Ice giants, form in a variety of surface patterns
    andamplitudes. It is thought that equatorial prograde surface
    zonal  flows are created by Reynolds-stress in rotation-dominated
    convection, whereas the equatorial retrograde jets form in an
    inertia-dominated regime due to turbulent angular momentum
    mixing. As far as numerical simulations can reach, simple
    Boussinesq-models predict only minor variation in the interior
    zonal flow structure along the axis of rotation according to the
    Taylor-Proudman-Theorem.
Further solar or stellar irradiation might cause a latitudinal
variation pattern of the surface temperature. The heat flux is then
expectably higher at the cooler poles and smaller at the hotter
equatorial region. If such a latitudinal
temperature/heat flux anomaly at the surface affects also the interior
of the fluid, thermal winds will introduce ageostrophic variations of
the geostrophic zonal flow structure along the axis of rotation.
To investigate the interaction between thermal wind driven
ageostrophic and Reynolds-stress driven geostrophic zonal flows, we
conduct a series of hydrodynamical simulations using a rotating and
convecting spherical fluid shell in the limit of the
Boussinesq-approximation. The fluid is heated by a homogeneously
distributed interior heat source and cooled by a
prescribed mean heat flux at the outer boundary, which additionally
varies smoothly with latitude as a axisymmetric spherical harmonic
degree two of variable amplitude.
Our results suggest, that for weak perturbation amplitudes the flow
tends to be the more ageostrophic the stronger the heat flux
normally. This results from increasing ageostrophic and decreasing
geostrophic zonal flows. Since the introduced thermal wind defines the
zonal flow variation along the axis of rotation, the equatoral jet
decreases its prograde amplitude. If the anomaly reaches a critical
strength, the equatorial jet crosses zero amplitude and can even be
reverted into retrograde direction. For perturbation amplitudes
stronger than critical, the reverted jet gains in amplitude due to
re-established,
but reverted Reynolds-stress. Even though the ageostrophic flow is
linearly amplified with the anomaly amplitude, the geostrophic flow
contributions dominate at maximal heat flux perturbation. We therefore
propose a nonlinear backreaction of the thermal wind onto the
Reynolds-stress driven zonal flows. Analysis of the main force balance
and parameter studies further foster this result.}
\conta{12:10 -- 12:25}
      {Michael Proctor}
      {}
      {Bounds and scaling for the Archontis dynamo}
{The Archontis dynamo is characterised by strong magnetic fields
closely aligned with the driving velocity field. An interesting
question is how far such a balance can
be maintained as diffusivity rates change. We present preliminary
results giving rigorous bounds on flow quantities and computations
showing that the Archontis balance is surprisingly robust.}
\end{mylist}          

\mysec{Lunch Break}{}
\vspace{0.2mm}
\begin{mylist}{19:00 -- 10:30 am}
     \item[12:25 -- 14:00]   Lunch is provided at ``One A The Square''.
       The restaurant is located at the North-West corner of the Main University Building. 
       Lunch vouchers are enclosed in your conference wallets.
\end{mylist}          

\myses{Session 2: Geophysical and Planetary Applications II}{Phil Livermore}
\begin{mylist}{19:00 -- 10:30 am}
      \conta{14:00 -- 14:15}
           {Graeme Sarson,}
           {R.~Boys, A.~Golightly and  D.~Henderson}
           {Modelling geomagnetic reversals as a
    Gaussian Cox Process}
{The mean rate of reversal of the geomagnetic field,
    as recorded in the Geomagnetic Polarity Time Scale (GPTS), has
    long been an object of study. The significance of the apparent
    long-term variations in the mean rate of reversals --- including
    the occurrence of superchrons --- has been debated, as has the
possible origins of these variations in external control of the
geodynamo (e.g.~by the time-varying boundary conditions imposed by
mantle convection). Here we model the long-term variations in the
reversal rate nonparametrically, in terms of an inhomogeneous Poisson
process. Specifically, we consider a Gaussian Cox process, a type of
doubly-stochastic Poisson process where the mean rate (or intensity)
is modelled in terms of a Gaussian process. Such processes are
amenable to likelihood-based inference using Bayesian Markov Chain
Monte-Carlo (MCMC) methods, which we employ to provide posterior
distributions of the model parameters. The specification of a Gaussian
process requires a covariance function, relating the intensity at
nearby times; crucially, however, the timescale of the covariance
function is not prescribed, but appears as a model hyperparameter,
whose posterior distribution is an important output of the
analysis. For the geodynamo, this hyperparameter should robustly
characterise the timescale of long-term variations. Two 
 different types of Gaussian Cox process are considered: a Log
 Gaussian Cox Process, applied to binned reversal data; and a
 Sigmoidal Gaussian Cox Process, applied to the discrete reversal data
 using a technique involving latent
variables. Different MCMC algorithms for sampling the posterior
distribution of the model parameters are investigated for both types
of process, to check (and to optimise) the convergence of the MCMC
chains. This analysis is applied to different records of the GPTS,
including those of Cande \& Kent (1995) and
Gradstein \& Ogg (1996). The implications of this analysis for the
geodynamo, and the possibility of comparable analysis of the output of
numerical geodynamo simulations, are discussed.
}
      \conta{14:15 -- 14:30}
           {Jack Wood}
           {}
           {A layer of fluid with variable conductivity}
{The electrical conductivity of Jupiter varies over
    several orders of magnitude from the non conducting outer reaches
    of the planet to the highly conducting metallic hydrogen region
    (at approximately 0.8 RJ). In an attempt to identify what effect
    this variance could have on the dynamics of such a system, I will
    present a plane layer approximation of a rapidly rotating,
    inviscid fluid with imposed zonal surface winds and varying
    conductivity. Limitations of a low $Rm$ assumption will be discussed
    along with initial thoughts for a higher $Rm$ approximation.
}
      \conta{14:30 -- 14:45}
           {Dali Kong,}
           {K.~Zhang and  G.~Schubert}
           {Gravitational signature of rotationally  distorted Jupiter caused by deep zonal winds}
{Both deep zonal winds, if they exist, and the basic
rotational distortion of Jupiter contribute to its zonal gravity
coefficients $J_n$ for $n$ greater than or equal to 2. In order to capture
the gravitational signature of Jupiter that is caused solely by its
deep
zonal winds, one must take into account the full effect of rotational
distortion by computing the coefficients $J_n$ in non-spherical
geometry.
Based on the model of a polytropic Jupiter with index unity, we
compute
for the first time Jupiter's gravity coefficients $J_2$, $J_4$, $J_6$, ...,
$J_{12}$
taking into account the full effect of rotational distortion of the
gaseous planet. For the model of deep zonal winds on cylinders
parallel
to the rotation axis, we also compute for the first time the variation
of
the gravity coefficients $\Delta J_2$, $\Delta J_4$, $\Delta J_6$, ..., $\Delta
J_{12}$ caused solely by the effect of the winds on the rotationally
distorted Jupiter. It is found that the effect of the zonal winds on
lower-order coefficients is weak, $|\Delta J_n/J_n|$ less than 1\%, for
$n=2,4,6$, but it is substantial for the high-degree coefficients with $n$
greater than or equal to 8.
}
\end{mylist}          

\myses{Posters and Afternoon Coffee Break}{Andrew Baggaley}
\begin{mylist}{19:00 -- 10:30 am}
      \item[14:45 -- 15:20] Posters
      \item[15:20 -- 15:50] Coffee break
\end{mylist}

\myses{Session 3: Solar Applications I}{Ineke De Moortel}
\begin{mylist}{19:00 -- 10:30 am}
      \conta{15:50 -- 16:10}
           {Thomas Neukirch,}
           {F.~Wilson, M.~Harrison}
           {One-dimensional force-free current sheet    equlibria: MHD
             vs Vlasov theory}
{One-dimensional current sheet equilibria are often
    used as initial states for investigations of fundamental plasma
    processes such as magnetic reconnection. While finding 1D MHD
    equilibria is very simple, the task of finding the corresponding
    equilibrium distribution functions in Vlasov theory is
    non-trivial. In this talk some recent results for constructing
    exact analytical equilibrium distribution functions for non-linear
    force-free current sheets will be presented.
}
      \conta{16:10 -- 16:25}
           {Frederick Gent,}
           {R.~Erd\'elyi, V.~Fedun and S.~Mumford}
           {Flux tube equilibrium}
{A single magnetic flux tube emerging from the solar
    photosphere to the lower corona is modelled in magnetohydrostatic
    equilibrium with a realistic stratified atmosphere. We solve
    analytically a 3D structure for the model, with field strength,
    plasma density, pressure and temperature all consistent with
    observational and theoretical estimates. Such flux tubes are
    observed to remain relatively stable for up to an hour or more,
    and it is our aim to apply the model as the background condition
    for numerical studies of energy transport mechanisms from the
    surface to the corona. The model includes a number of free
    parameters, which makes the solution applicable to a variety of
    other physical problems and it may therefore be of more general
    interest.
}
      \conta{16:25 -- 16:40}
           {Eduard Kontar,}
           {N. Bian, I. Hannah and N.Jeffrey}
           {Magnetic fluctuations, acceleration and    transport in
             solar flares}
{Plasma turbulence is thought to be associated with
    various physical processes involved in solar flares, including
    magnetic reconnection, particle acceleration, and transport. Using
    X-ray observations, we determine the spatial and spectral
    distributions of energetic electrons for a flare, which was
    previously found to be consistent with a reconnection
    scenario. Energy-dependent transport of tens of keV electrons is
    observed to occur both along and across the guiding magnetic field
    of the loop. We show that the cross-field transport is consistent
    with the presence of magnetic turbulence in the loop, where
    electrons are accelerated, and estimate the magnitude of the
    magnetic field fluctuations. The level of magnetic fluctuations
    peaks when the largest number of electrons is accelerated and is
    below detectability or absent at the decay phase.}
      \conta{16:40 -- 16:55}
           {Nic Bian,}
           {and E.~Kontar}
           {Stochastic acceleration by multi-island
    contraction during turbulent magnetic reconnection}
{The acceleration of charged particles in magnetized
    plasmas is considered during turbulent multi-island magnetic
    reconnection. The particle acceleration model is constructed for
    an ensemble of islands which produce adiabatic compression of the
    particles. The model takes into account the statistical
    fluctuations in the compression rate experienced by the particles
    during their transport in the acceleration region. The evolution
    of the particle distribution function is described as a
    simultaneous first- and second-order Fermi acceleration
    process. While the efficiency of the first-order process is
    controlled by the average rate of compression, the second-order
    process involves the variance in the compression rate. Moreover,
    the acceleration efficiency associated with the second-order
    process involves both the Eulerian properties of the compression
    field and the Lagrangian properties of the particles. The
    stochastic contribution to the acceleration is nonresona
 nt and can dominate the systematic part in the case of a large
 variance in the compression rate. The model addresses the role of the
 second-order process, how the latter can be related to the
 large-scale turbulent transport of particles, and explains some
 features of the numerical simulations of particle acceleration by
 multi-island contraction during magnetic reconnection.
 \url{http://adsabs.harvard.edu/abs/2013PhRvL.110o1101B}
}
      \conta{16:55 -- 17:10}
           {David Pontin,}
           {A.L.~Wilmot-Smith and G.~Hornig}
           {On the relaxation of braided magnetic    fields}
{The braiding of the solar coronal magnetic field by
    photospheric motions - with subsequent relaxation and plasma
    heating - is one of the most widely debated ideas of solar
    physics. We describe here the resistive relaxation of a magnetic
    field that contains braided magnetic flux, as a model for a
    coronal loop, subject to line-tied boundary conditions. It is
    found that the system is unstable, and that following the onset of
    instability a myriad of thin current layers progressively
    develops. The current layer structure becomes increasingly
    complex, and is increasingly long-lived, for higher magnetic
    Reynolds numbers ($Rm$). For large $Rm$ the evolution resembles a
    state of decaying turbulence. The final state is found to
    approximate a non-linear force-free field, implying that the
    system does not undergo a 'Taylor-like' relaxation. The
    implications of the results for the heating of the solar corona
    will be discussed.}
      \cont{17:10 -- 17:15} {Keke Zhang}{} {Brief announcement}
\end{mylist}

\mysec{Conference Dinner at The Bothy}{}
\vspace{0.2mm}
\begin{mylist}{19:00 -- 10:30 am}
    \item[18:30 -- 19:20 ]          Drinks at the Maths \& Stats Common Room
    \item[19:30 -- ]           Dinner at The Bothy
\end{mylist}          

\subsubsection*{--- Friday, 2014-05-23 ---}{}

\myses{Invited Lecture}{David Hughes}
\begin{mylist}{19:00 -- 10:30 am}
      \conta{9:00 -- 9:40}
          {Chris A.~Jones}
          {} 
          {Dynamo models of Jupiter's magnetic field}
{ Numerical dynamo models have had some success in reproducing
 important features of the Earth's magnetic field. Here we report
 on simulations of Jupiter's magnetic field using the anelastic
 approximation, which takes into account the large density variation
 across the dynamo region. The reference state used in these
 models is a Jupiter model taken from ab initio calculations of the
 physical properties of Jupiter's magnetic field (French et al. 2012),
 which makes the reasonable assumption that the interior is close to
 adiabatic. The French et al. work also gives an electrical
 conductivity
 profile which is adopted here.
 Dynamo simulations depend on the dimensionless input parameters,
 partibularly the Ekman number, Rayleigh number, the Prandtl
 number and magnetic Prandtl number. Many different types of field
 have been found, some of which will be described. The most relevant
 models are those which produce a Jupiter-like strong dipole dominated
 field. These are found at low Ekman number, Rayleigh numbers large
 enough for the convective heat flux to dominate the radiative flux,
 low
 Prandtl number and moderate magnetic Prandtl number. Another
 important
 issue is the driving heat flux source. Here we assume that Jupiter
 evolves through a sequence of adiabats, leading to a distributed
 entropy source throughout the planet, rather than basal heating from
 the
 small rocky core. The interaction between the magnetic field, the
 zonal
 flow and the convection appears to be crucial in determining the type
 of magnetic field found.}
\end{mylist}          

\myses{Session 4: Solar Applications II}{David~Tsiklauri}
\begin{mylist}{19:00 -- 10:30 am}
      \conta{9:40 -- 9:55}
           {Anvar Shukurov,}
           {P.~Bushby, L.~Cole and Y.~Ji}
           {Asymptotics for the mean-field dynamo in a
    slab}
{Boundary-layer asymptotics for the kinematic
    mean-field dynamo equations in a slab are revisited, earlier
    results revised, and some misconceptions identified to achieve a
    consistent understanding of the dynamo action in a broad range of
    dynamo numbers.}
      \conta{9:55 -- 10:10}
           {Andrew Gascoyne}
           {and R. Jain}
           {p-Mode driven sausage waves}
{We investigate the generation and propagation of
    sausage tube waves within the solar convection zone and
    chromosphere by the buffeting of p modes.  The tube waves
    propagate along the many magnetic fibrils which are embedded in
    the convection zone and expand into the chromosphere due to the
    fall in density with height of the surrounding plasma.  The
    magnetic fibrils form a waveguide for these waves to freely
    propagate up and down the tube, those waves propagating upward
    pass through the photosphere into the chromosphere and enter the
    upper atmosphere, where they can be measured as loop oscillations
    and other forms of propagating coronal waves.  We treat the
    magnetic fibrils as vertically aligned, thin flux tubes embedded
    in a two layer polytropic-isothermal atmosphere to investigate the
    coupling of p-mode driven sausage waves; which are excited in the
    convection zone and propagate into the overlying chromosphere.
    The excited tube waves carry energy away from
  the p-mode cavity resulting in a deficit of p-mode energy which we
  quantify by computing the associated damping rate of the driving p
  modes.  We calculate the damping rates and compare them with
  observations and previous theoretical studies of this nature.
}
      \conta{10:10 -- 10:25}
           {Peter~Wyper}
           {and R.~Jain}
           {The effects of current sheet asymmetry on 3D null
             reconnection}
{Asymmetric current sheets are prevalent in both
    astrophysical and laboratory plasmas with complex three
    dimensional (3D) magnetic topologies. In 3D, reconnection rate is
    measured by $[ \int{ E_{\parallel} dl} ]_{max}$  along all field
    lines threading
    the non-ideal region. Therefore, current sheet dimensional
    asymmetry likely plays an important role in the manner and rate of
    3D reconnection. This work studies the importance of current sheet
    asymmetry for the
fan and spine modes of 3D null point reconnection.  We find that
asymmetric fan reconnection is characterised by an asymmetric
reconnection of flux past each spine line and a bulk flow of plasma
through the null point. In contrast, asymmetric spine reconnection is
inherently equal and opposite in how flux is reconnected across the
fan plane.  The rich nature of the spine solutions necessitates in
some circumstances a dual definition for the reconnection rate: a
local rate quantifying the ``total'' rate that flux crosses the fan
plane and a global rate quantifying the rate an observer ``sees'' flux
transferred far from the null. Both models suggest that current sheet
asymmetry has a profound effect on 3D null reconnection.
}
      \conta{10:25 -- 10:40}
           {Christina Davies}
           {}
           {Short wavelength magnetic buoyancy instability}
 {We consider the magnetic buoyancy instability in the
   short-wavelength limit of Gilman (1970). In this limit the
   perturbation equations (a  system of coupled ODEs) can be reduced
   to a single algebraic dispersion relation, with coefficients
   depending on height. Put otherwise it seems that, in this limit, a
   problem that would have been  treated as an eigenvalue problem
   requiring a set of boundary conditions can be reduced to a single
   equation for which the boundary conditions are unimportant. Here I
   present asymptotics and numerical calculations to illustrate the
   link between the two systems, which can be viewed as being
   analogous to the more familiar problem of  the quantum harmonic
   oscillator.} 
\end{mylist}          

\mysec{Morning Coffee Break}{}
\vspace{2mm}
\begin{mylist}{19:00 -- 10:30 am}
      \item[10:40 -- 11:10]
\end{mylist}          

\myses{Session 5:  Solar Applications III}{Paul Bushby}
\begin{mylist}{19:00 -- 10:30 am}
      \conta{11:10 -- 11:25}
           {Ineke De Moortel,}
           {D.J.~Pascoe, A.W.~Hood, A.N.~Wright,
    M.S.~Ruderman and J.~Terradas}
           {Wave damping due to mode coupling in Solar
    coronal loops}
{Recent observations of the loops in the Sun's
    atmosphere reveal ubiquitous transverse velocity perturbations
    that undergo strong damping as they propagate. Using numerical
    (and analytical) modelling, we demonstrate that these can be
    understood in terms of coupling of different wave modes in the
    inhomogeneous boundaries of the loops:  we perform 3D numerical
    MHD simulations of footpoint-driven transverse waves propagating
    in a coronal plasma with a cylindrical density structure. Mode
    coupling in the inhomogeneous boundary layers of the loops leads
    to the coupling of the transversal (kink) mode to the azimuthal
    (Alfven) mode, observed as the decay of the transverse kink
    oscillations.
Both the numerical and analytical results show that the initial
damping is Gaussian in nature, before tending to linear exponential
damping at large heights. To use the observed waves as a seismological
tool, we propose a general spatial damping profile (based on the
analytical results) that accounts for the initial Gaussian stage of
damped kink waves as well as the asymptotic exponential stage. The
applicability of this profile for coronal seismology purposes is
demonstrated by a full numerical parametric study of the relevant
physical parameters.
}
      \conta{11:25 -- 11:40}
           {David~Tsiklauri}
           {}
           {3D, linearly polarised, Alfven wave dissipation in
             Arnold-Beltrami-Childress magnetic field}
{Previous studies (e.g. Malara et al ApJ, 533, 523 (2000)) considered
small-amplitude Alfvenic wave (AW) packets in WKB approximation. They
draw a
distinction between 2D AW dissipation via phase mixing and 3D AW
dissipation via exponentially divergent magnetic field lines. In the
former case AW dissipation
time scales as $S^{1/3}$ and in the latter as $\log(S)$, where $S$ is the
Lundquist number. In this work, using MHD numerical simulation and
Arnold-Beltrami-Childress
magnetic field, we verify the log(S) scaling via direct MHD numerical
simulation for the first time. Implications for the MHD wave heating
of solar corona are
also discussed.}
      \conta{11:40 -- 11:55}
           {Samuel~Hunter,}
           {D.~Hughes and S.~Griffiths}
           {Waves in shallow water MHD}
{Models of the solar dynamo predict the presence of a strong magnetic
field permeating this structure. Effects of fluid motion in the
tachocline occur on all time--scales: from the short, where
perturbations from convective overshooting displace fluid; through the
medium, in which the sun's 11--year cycle takes place; to the long,
where solar structure is determined.
Here we adopt the shallow water MHD equations [Gilman, 2000], which
are an extension of the well--studied hydrodynamic shallow water
system to incorporate the extra effects of magnetised fluid. Waves are
a fundamental property of fluid systems, and such shallow
astrophysical layers almost certainly exhibit wave--like
behaviour. Two and three layer models have been used extensively to
analyse wave motion in the ocean. Here we augment this system to
describe waves in the presence of a magnetic field; magnetically
modified internal and surface modes result, that are combinations of
Alfv\'en and gravity waves.}
      \conta{11:55 -- 12:10}
           {Michael~Bareford}
           {and A.~Hood}
           {Shock heating in the Solar corona}
{
We explore the process by which a coronal loop can
    become heated in response to an ideal
magnetic field instability. A three-dimensional magnetohydrodynamic
Lagrangian-remap code
is used to simulate the evolution of a specific line-tied field
configuration, which is based on a
zero-net-current cylindrical loop model. The initial loop state is
known to be linearly kink unstable.
In addition, the field surrounding the loop is potential (the external
field is parallel to the initial loop axis).
The kink instability rapidly leads to the formation of slow mode
shocks within the loop interior, where the pressure and sound speed
are low compared to the loop boundary. We investigate how these
slow mode shocks influence the heating process. In general, slow mode
shocks act to release magnetic  energy in the form of currents, which
then, via the Lorentz force, concentrate plasma flows, giving rise to
steep velocity gradients at the loop boundary. It is this last feature
that causes shock heating, which is represented in the code as an
artificial viscosity. Our model also incorporates
thermal conduction, radiation and gravity; thus, we can forward model
our results to show the appearance of the loop within the 171 A
passband used by the AIA instrument onboard the Solar Dynamics
Observatory.
}
      \conta{12:10 -- 12:25}
           {David~MacTaggart}
           {}
           {Finite deformation in ideal MHD: analytical
    twisted current layers}
{
Numerical studies of magnetic relaxation take, as
    their initial conditions, stressed magnetic fields out of
    equilibrium. To produce the required initial condition, flows are
    often imposed on the boundaries of the computational domain to
    deform a simple magnetic field into the required shape. In this
    talk we shall look at performing such nonlinear deformations
    analytically. We shall examine how to deform the magnetic field
    and the current density and demonstrate this with the example of
    deforming a straight magnetic field into one that produces twisted
    current layers. The technique can be applied to other magnetic
    fields, e.g. null points, and will prove useful to modellers in
    setting up initial conditions for relaxation problems.
}
\end{mylist}          

\mysec{Lunch Break}{}
\vspace{0.2mm}
\begin{mylist}{19:00 -- 10:30 am}
     \item[12:25 -- 14:00]   Lunch is provided at ``One A The Square''.
       The restaurant is located at the North-West corner of the Main
       University Building. 
       Lunch vouchers are enclosed in your conference wallets.
\end{mylist}

\myses{Session 6: Models and Methods}{Graeme Sarson}
\begin{mylist}{19:00 -- 10:30 am}
      \conta{14:00 -- 14:15}
           {Stuart Mumford}
           {}
           {3D Simulations of MHD waves in low solar atmospheric flux
             tubes driven by photospheric motions}
{
Recent ground- and space-based observational results reveal the
presence of small-scale motion between convection cells in the solar
photosphere.In these regions small-scale magnetic flux tubes are
generated due to the interaction of granulation motion and background
magnetic field. In this talk we present numerical simulations of
magnetic flux tubes in these regions driven by photospheric motion.
A new method for analysing MHD wave modes in 3D simulations is
presented which decomposes the velocity perturbations into three
components defined as parallel to the magnetic field, perpendicular to
the
magnetic field and flux surface and an azimuthal component
perpendicular to the field inside the surface. Allowing the accurate
identification of MHD modes in complex 3D geometries.
 This method is implemented and applied to the numerical simulations
and then used to identify excited wave modes and the energy flux of
each is calculated. A comparison of the excited wave modes and the
energy flux contributed by each for a uniform torsional driver, a
Archimedean sprial-type and a logarithmic sprial-type drivers with
each other and with vertical and horizontal drivers is presented here.
}
      \conta{14:15 -- 14:30}
           {Wayne~Arter}
           {}
           {Potential vorticity formulation of
    compressible Magnetohydrodynamics}
{Compressible ideal magnetohydrodynamics (MHD) is
    formulated in terms of the
time evolution of potential vorticity and magnetic flux per unit mass
using a compact Lie bracket notation.
It is demonstrated that this simplifies analytic solution
in at least one very important situation relevant to magnetic fusion
experiments. Potentially important implications
for analytic and numerical modelling of both laboratory
and astrophysical plasmas are also discussed.
}
      \conta{14:30 -- 14:45}
           {Ashley Willis}
           {}
           {Optimising the dynamo}
{ What velocity field is `best' for generating a
    magnetic field?  What is the lowest possible magnetic Reynolds
    number for a magnetic dynamo?  To answer such fundamental
    questions requires optimisation over the full space of permitted
    velocity fields.  Theoretical studies have considered many
    parametrized velocity fields in several geometries, including
    Ponomorenko, Roberts, Arnold-Beltrami-Childress (ABC), and Dudley
    and James flows.
In this work, optimisation is shown to be possible without need for
the specification of a parametrized set
of acceptable flows [1].  This enables a lower bound on the magnetic
Reynolds number to be identified for a dynamo, here for the case of a
periodic box.
[1] A. P. Willis (2012), PRL, 109, 251101.
}
      \conta{14:45 -- 15:00}
           {Michael~Griffiths,}
           {V.~Fedun and R.~Erdelyi}
           {A fast MHD code for gravitationally
    stratified media using multiple graphical processing units: SMAUG}
{
Parallelisation techniques have been exploited most
    successfully by the gaming/graphics industry with the adoption of
    graphical processing units (GPUs) possessing hundreds of processor
    cores. The opportunity has been recognised by the computational
    sciences and engineering communities who have recently harnessed
    extremely well the numerical performance of GPUs. For example,
    parallel magnetohydrodynamic (MHD) algorithms are important for
    numerical modelling of highly inhomogeneous solar, geophysical and
    astrophysical plasmas. Here we describe the implementation of the
    new GPU-based MHD code, the Sheffield Magnetohydrodynamics
    Algorithm Using GPUs (SMAUG). SMAUG is a 3D nonliner MHD code
    capable of modelling magnetised and gravitationally strongly
    stratified plasmas.
The objective of this presentation is to introduce the numerical
methods used and the techniques for porting the code to this novel and
highly parallel compute architecture. We show the recent development
enabling the MHD code to utilise multiple GPUs and we describe the
techniques implemented to enable communication between the GPUs.  The
methods applied are justified by the performance benchmarks and
validation results demonstrating that the code successfully simulates
the physics for a range of test scenarios including a full 3D
realistic model of MHD wave propagation in the strongly stratified
solar atmosphere.
}
      \conta{15:00 -- 15:15}
           {Kacper~Kornet}
           {and A.~Potherat}
           {A new spectral method for direct numerical simulations of
             magnetohydrodynamic channel flows}
{
Simulations of liquid metal flows in channel and
    duct configurations under a strong magnetic field pose a difficult
    problem for existing numerical methods.  The main obstacle is the
    linear increase in number of modes required to resolve thin
    Hartmann boundary layers with the intensity of the magnetic field
    $B$. To overcome this problem  we developed a new approach to the
    numerical calculations describing these flows. The solution of the
    flow is expressed in a base of eigenfunctions of the linear part
    of the governing equations and its adjoint.   We show that in this
    approach the computational cost does not depend on the thickness
    of boundary layer and therefore it allows for performing
    calculations for high magnetic fields.
}
\end{mylist}          

\mysec{End of Meeting Coffee Break}{}
\vspace{2mm}
\begin{mylist}{19:00 -- 10:30 am}
      \item[15:15 -- 16:00]
\end{mylist}

\mysec{School of Mathematics and Statistics Colloquium}{}
\begin{mylist}{19:00 -- 10:30 am}
      \conta{16:00 -- 17:00}
           {Prof Herbert Huppert, FRS}
           {}
           {Fluid modelling of carbon dioxide sequestration \\
{\small \textsl{Note}: This is a School event. UK MHD Participants are most welcome to
attend. The Colloquium takes place in the lecture theatre next door - Maths
516.}}
{Current global anthropogenic emissions of carbon dioxide are
  approximately 32 Gigatonnes annually. The influence of this
  green-house gas on climate has raised concern. A means of reducing
  environmental damage is to store carbon dioxide somewhere until well
past the end of the fossil fuel era. Storage by injection of liquid,
or supercritical, carbon dioxide into porous reservoir rocks, such as
depleted oil and gas fields and regional saline aquifers, is being
considered. The presentation will discuss the rate and form of
propagation to be expected and quantify some of the risks
involved. The talk builds on theoretical and experimental
investigations of input of liquid of one viscosity and density from a
point source above an impermeable boundary, either horizontal or
slanted, into a heterogeneous porous medium saturated with liquid of
different viscosity and density. In the Sleipner natural gas field,
carbon dioxide has been injected at a rate of ~ 1 Mt/yr since 1996. We
will briefly show how to apply our results to interpret these field
observations. One of the best controlled field experiments, the Otway
Project, commenced on 2 April 2008 in Victoria,
Australia. Approximately sixty thousand tonnes of carbon dioxide was
injected into a slanted sill over a period of just over a year. We
will show how accurately some of our theoretical models predict the
field data obtained so far. The talk will be illustrated by colour
movie sequences of laboratory experiments and some simple desk-top
demonstrations of aspects of flow of multi-phase fluids into a porous ambient.}
\end{mylist}

\subsubsection*{--- Posters ---}{}
Posters will remain on display in the Maths and Stats Common Room on
both days.\vspace{2mm}
\begin{mylist}{19:00 -- 10:30 am}
      \conta{Poster}
           {Homayon Aryan,}
           {M.A.~Balikhin, A.~Taktakishvili and T.L.~Zhang}
           {CME associated shocks}
{The interaction of CMEs with the solar wind can lead
    to the formation of interplanetary shocks. Ions accelerated at
    these shocks contribute to the solar energetic protons observed in
    the vicinity of the Earth. Recently a joint analysis of Venus
    Express (VEX) and STEREO data by C. T. Russell and co-authors have
    shown that the formation of strong shocks associated with
    Co-rotating Interaction Regions (CIRs) takes place between the
    orbits of Venus and the Earth as a result of coalescence of weaker
    shocks formed earlier. The present study uses VEX and Advanced
    Composition Explorer (ACE) data in order to analyse shocks
    associated with CMEs that occurred on 29th and 30th of July 2007
    during the solar wind conjunction period between Venus and the
    Earth. For these particular cases it is shown that the above
    scenario of shock formation proposed for CIRs also takes place for
    CMEs. Contradiction with shock formation resulting from MHD
    modelling is explained by inability of classical MHD to account
    for the role of wave dispersion in the formation of the shock.
}
      \conta{Poster}
           {Jonathan Hodgson}
           {}
           {A formalism for computing symmetric MHD
    equilibria with anisotropic pressure}
{We present a general formalism that allows the
    calculation of MHD equilibria with translational or rotational
    symmetry and anisotropic pressure. The pressure tensor is assumed
    to be of CGL form, and external forces such as gravity or
    centrifugal forces are included. If field-aligned currents are
    assumed to vanish, the fundamental equation is of Grad-Shafranov
    type. The inclusion of field-aligned currents results in a more
    complex mathematical problem with no obvious analogy in the theory
    of MHD equilibrium with isotropic pressure. Magnetospheres of
    giant, fast rotating planets could be a possible application of
    this theoretical formalism.
}
      \conta{Poster}
           {Agnieszka Hudoba,}
           {J.~Priede, S.~Aleksandrova and S.~Molokov}
           {Instabilities in the buoyant convective
    flows subject to high magnetic fields}
{The project encapsulates the study of instabilities
    in the buoyant convective flows in the presence of high magnetic
    fields. The study is of particular importance as the phenomenon
    plays a fundamental role in such industrial applications as fusion
    reactor blankets, semiconductor crystal growth or electromagnetic
    processing of materials. Buoyant magnetoconvection has been
    studied numerically involving spectral methods. The linear
    stability results show that the instability critically depends on
    the electrical and thermal boundary conditions, and on the Prandtl
    and Hartmann numbers. The problem is first solved in the
    inductionless approximation, where the flow cannot disturb the
    imposed magnetic field. Then the higher magnetic Reynolds number
    is considered in order to study the magnetic field evolution and
    possible Alfvenic oscillations.
}
      \conta{Poster}
           {David Skinner}
           {}
           {Double-diffusive magnetic buoyancy
    instability in a quasi-two-dimensional Cartesian geometry}
{Magnetic buoyancy, believed to occur in the solar
    tachocline, is both an important part
of large-scale solar dynamo models and the picture of how sunspots are
formed. Given
that in the tachocline region the ratio of magnetic diffusivity to
thermal diffusivity
is small it is important, for both the dynamo and sunspot formation
pictures, to
understand magnetic buoyancy in this regime. Furthermore, the
tachocline is a region
of strong shear and such investigations must involve structures that
become buoyant
in the double-diffusive regime which are generated entirely from a
shear flow.
Silvers et al. (2009) have illustrated that shear-generated
double-diffusive magnetic
buoyancy instability is possible in the tachocline. However, this
study was severely limited due to the computational requirements of
running three
dimensional magnetohydrodynamics simulations over diffusive
timescales. A more
comprehensive investigation
is required to fully understand the double-diffusive magnetic buoyancy
instability and
its dependency on a number of key parameters; such an investigation
requires the
consideration of a reduced model.
Here we consider a quasi-two dimensional model where all gradients in
the $x$ direction are set to zero. We show how the instability is
sensitive to
changes in the
thermal diffusivity and also show how different initial configurations
of the forced shear
flow affect the behaviour of the instability. Finally, we conclude
that if the tachocline is
thinner than currently stated then the double-diffusive magnetic
buoyancy instability
can more easily occur.
}
      \conta{Poster}
           {Christopher Davies,}
           {L.~Silva and J.~Mound}
           {Modelling the influence of a translating
    inner core on outer core convection}
{It has recently been proposed that the hemispheric
    seismic structure of the inner core
can be explained by a self-sustained rigid-body translation of the
inner core material,
resulting in melting of the solid at the leading face and a
compensating crystallisation
at the trailing face. This process induces a hemispherical variation
in the release of light
elements and latent heat at the inner-core boundary, the two main
sources of thermochemical
buoyancy thought to drive convection in the outer core, suggesting a
possible influence of
inner core translation on outer core dynamics. The outer core is also
likely subjected to
lateral variations in heat-flux imposed from above by the convecting
mantle. We use numerical
models of nonmagnetic thermal convection in a rotating spherical shell
to investigate possible
long-term effects of inner core translation and mantle convection on
the outer core.
}
      \conta{Poster}
           {Jonathan Hagan}
           {}
           {Two-dimensional nonlinear travelling waves in MHD channel flow}
{
The present study is concerned with the stability of a flow of viscous
conducting liquid driven by a pressure gradient between two parallel walls in the
presence of a transverse magnetic field. This magnetohydrodynamic
counterpart of the classic plane Poiseuille flow is generally known as
the Hartmann flow. Although the magnetic field has a strong
stabilizing effect, the turbulence is known to set in this flow
similarly to its hydrodynamic counterpart well below the threshold
predicted by the linear stability theory. Such a nonlinear transition
to turbulence is thought to be mediated by unstable equilibrium flow
states which may exist in addition to the base flow. In the plane
Poiseuille flow, such 
states are known to bifurcate from the base flow as 2D travelling
waves that appear due to the subcritical instability at the critical
Reynolds number $Re_c =5772$ and exist down
to $Re_2 = 2939$. The aim of the present study is to investigate how
these subcritical states
are affected by the magnetic field. We find that in the Hartmann flow,
2D nonlinear
travelling waves exist starting from the linear stability threshold
$Re_c \approx 4.83 \times 10^4 Ha$ down
to $Re_2 \approx 6.5\times10^3 Ha$, where Ha is the Hartmann number. Note that $Re_2$ is
still by an order
of magnitude greater than the experimentally observed value for the
onset of turbulence
in this flow. Three-dimensional travelling waves bifurcating either
from 2D ones or
infinity are expected to extend beyond Re2 and give a more adequate
prediction for the
onset of turbulence.
}
\end{mylist}          

\subsubsection*{--- Withdrawn Contributions---}{}
A small number of contributions were withdrawn for valid
reasons. Their details are included below.
\begin{mylist}{19:00 -- 10:30 am}
      \conta{Withdrawn talk}
           {Aditi~Sood}
           {and E.-J.~Kim}
           {Dynamic model of dynamo (magnetic activity)
    and rotation}
{A dynamic model of dynamo and rotation is
    investigated to understand the observational data of the
    dependence
of the magnetic activities and the differential rotation
$\Delta\Omega$ on the rotation rate $\Omega$.
Specifically, we propose a minimal seventh order non-linear dynamical
system for magnetic fields and differential rotation $\Delta\Omega$
by parameterizing the generation and destruction of magnetic fields by
$\alpha$-$\Omega$ effect and magnetic flux loss from stars and by
including quenching of $\alpha$ effect and differential rotation
$\Delta\Omega$ due to  Lorentz force. By examining different forms of
alpha
quenching and flux loss,  we study how the strength and frequency
$\omega$ of magnetic fields, and the differential rotation
$\Delta\Omega$
change with the rotation rate $\Omega$ through dynamo number. In
particular, among the three cases with (i) $\alpha$-quenching and no
flux
loss (ii) flux loss and no $\alpha$-quenching; (iii)
$\alpha$-quenching and flux loss,
our results show that the best agreement with observations is obtained
in the case (iii) with equal amount of alpha quenching
and poloidal and toroidal magnetic flux losses with quadratic
nonlinear dependence on $|B|$. Specifically, in this case,
the frequency spectrum of magnetic field has a well-localized
frequency of the maximum intensity which scales
as $\omega\propto\Omega^{0.80}$, in agreement with observation of
Noyes et al. (1984), while
magnetic field and mean differential rotation exhibit the tendency of
saturation for high rotation.
Implication of our results in light of necessary dynamic balance is
discussed. Detailed investigations leads us to conclusions that
dynamic balance between $\alpha$ source term and destruction of
magnetic field as well as various transport coefficients
is necessary for a  stellar dynamo to operate near marginal stability
which could be result of self organizing process.
}
\conta{Withdrawn talk}
    {Grace Cox}
    {P.~Livermore and J.~Mound}
    {Forward models of torsional waves:  dispersion and geometric effects}
    {Alfv\'{e}n waves are a set of transverse waves that
    propagate in an electrically conducting fluid in the presence of
    an ambient magnetic field. Studies of such waves in the Earth's
    interior are important because inverse methods allow us to make
    inferences about the structure and physical properties of the core
    that would otherwise remain inaccessible. We produce 1D forward
    models of cylindrical torsional Alfv\'{e}n waves in the Earth's
    core, also known as torsional oscillations, and study their
    evolution in a full sphere and an equatorially symmetric spherical
    shell. Here we find that  travelling torsional waves undergo
    significant geometric dispersion that increases with successive
    reflections from the boundaries such that an initial wave pulse
    becomes unidentifiable within three transits of the core. We
    investigate the relationship between geometric dispersion and
    wavelength, concluding that long wavelength features are more
    dispersive than short wavelength features. This result is particularly important because torsional
 waves that have been inferred in the Earth's core from secular
 variation are relatively long wavelength, and are therefore likely to
 undergo significant dispersion within the core. When stress free
 boundary conditions on angular velocity are applied, waves are
 reflected at the equator of the core-mantle boundary ($s=r_c$) with
 the same sign as the incident wave. Waves that pass through the
 rotation axis undergo a pseudo-reflection and display more
 complicated behaviour due to a phase shift. In an equatorially
 symmetric shell, we identify a weak reflection at the tangent
 cylinder due to geometric effects.
}

\end{mylist}

\section*{--- List of Participants ---}{}
\begin{multicols}{2}
\begin{enumerate}[itemsep=-1pt,parsep=2pt]
\item  Michael Proctor, University of Cambridge
\item  Andrew M. Soward, University of Exeter
\item  Radostin Simitev, University of Glasgow
\item  Andrew Baggaley, University of Glasgow
\item  David Fearn, University of Glasgow
\item  J B Taylor, Radwin, Wallingford
\item  Lintao Zhang, Coventry University
\item  Kelig Aujogue, Coventry University
\item  Kacper Kornet, Coventry University
\item  Ivan Pakhotin,  University of Sheffield
\item  Homayon Aryan, University of Sheffield
\item  Peter Wyper, University of Sheffield
\item  Dali Kong, University of Exeter
\item  Keke Zhang, University of Exeter
\item  Andrew Fletcher,  Newcastle University
\item  Mitchell Berger, University of Exeter
\item  Wayne Arter, CCFE, Abingdon
\item  Samuel Hunter, Univesity of Leeds 
\item  Michael Griffiths, University of Sheffield
\item  Michael Bareford, University of St Andrews
\item  Phil Livermore,  University of Leeds
\item  David MacTaggart, University of Abertay Dundee
\item  David Tsiklauri,  Queen Mary University of London
\item  Stuart Mumford, University of Sheffield
\item  Lara Silvers, City University London
\item  William Brown, University of Leeds
\item  David Hughes, University of Leeds
\item  Christina Davies, University of Leeds
\item  Frederick Gent, University of Sheffield
\item  Robert Teed, University of Leeds
\item  Paul Bushby, Newcastle University
\item  Moritz Linkmann,  University of Edinburgn
\item  Lee Braiden, Coventry University
\item  Grace Cox, University of Leeds
\item  Laura Burgess, University of Leeds
\item  Jonathan Hodgson, University of St Andrews
\item  Wieland Dietrich, University of Leeds
\item  Chris Jones, University of Leeds
\item  David Skinner, City University London
\item  Thomas Neukirch, University of St Andrews
\item  Agnieszka Hudoba, Coventry University
\item  Jack Wood, University of Leeds
\item  Andrew Gascoyne, University of Sheffield
\item  Ineke De Moortel, University of St Andrews
\item  Sam Durston, University of Exeter
\item  Luis Silva, University of Glasgow
\item  Benjamin Favier, University of Cambridge
\item  Christopher Davies, University of Leeds
\item  Graeme Sarson, Newcastle University
\item  Anvar Shukurov, Newcastle University
\item  David Pontin, University of Dundee
\item  Jonathan Hagan, Coventry University
\item  Eduard Kontar, University of Glasgow
\item  Nic Bian, University of Glasgow
\item  Natasha Jeffrey,  University of Glasgow
\item  Philip McGavin, University of Dundee
\item  Iain Hannah, University of Glasgow
\item  Hugh Hudson, University of Glasgow
\item  Ashley Willis, University of Sheffield
\item  Scott Richardson, University of Dundee
\item  Alasdair Wilson, University of Glasgow
\item  Heather Ratcliffe, University of Glasgow
\end{enumerate}
\end{multicols}

\end{document}